\documentstyle[11pt,epsfig]{article}
\def\be {\begin{equation}}
\def\ee {\end{equation}}
\def\bea {\begin{eqnarray}}
\def\eea {\end{eqnarray}}
\def\bc {\begin{center}}
\def\ec {\end{center}}
\def\bfg {\begin{figure}}
\def\efg {\end{figure}}
\def\bi {\begin{itemize}}
\def\ei {\end{itemize}}

\def\pa {\partial}

\title{Weyl Anomaly and Initial Singularity Crossing}
\author{Adel Awad  \\
Centre for Theoretical Physics,
British University of Egypt\\
Sherouk City 11837, Egypt\\
and\\
Department of Physics, Faculty of Science,
Ain Shams University\\ Cairo 11566, Egypt\\
}
\date{}
\begin{document}

\maketitle

\begin{abstract}

 We consider the role of quantum effects, mainly, Weyl anomaly in modifying FLRW model singular behavior at early times. Weyl anomaly corrections to FLRW models have been considered in the past, here we reconsider this model and show the following: The singularity of this model is weak according to Tipler and Krolak, therefore, the spacetime might admit a geodesic extension. Weyl anomaly corrections changes the nature of the initial singularity from a big bang singularity to a sudden singularity. The two branches of solutions consistent with the semiclassical treatment form a disconnected manifold. Joining these two parts at the singularity provides us with a $C^1$ extension to nonspacelike geodesics and leaves the spacetime geodesically complete. Using Gauss-Codazzi equations one can derive generalized junction conditions for this higher-derivative gravity. The extended spacetime obeys Friedmann and Raychaudhuri equations and the junction conditions. The junction does not generate Dirac delta functions in matter sources which keeps the equation of state unchanged. \\

\end{abstract}
\section{Introduction}
\vspace*{0.5cm}

One of the main results concerning spacetime singularities is the known collection of singularity theorems due to Penrose and Hawking \cite{P-H}. These theorems prove the existence of singularities for a general class of spacetimes with certain energy conditions and global properties. A classic example of such spacetimes is that describing Friedmann-Lemaitre-Robertson-Walker (FLRW) models with initial or big bang singularity.

An important task of any possible quantum gravity theory is to resolve spacetime singularities of General Relativity. Until we have such a full resolution, it is constructive to ask how large is the impact of quantizing matter sources on spacetime singularities. We consider the role of quantum effects, mainly, Weyl/Trace anomaly on modifying FLRW model singular behavior at early times. Weyl anomaly \cite{Duff} is one of the interesting phenomena in quantum field theory on curved spaces that arises at one-loop level for a collection of conformal fields. It is known that quantum corrections due to Weyl anomaly are geometric in nature and takes the form of higher-derivatives terms added to Einstein field equations. Here we only consider Weyl anomaly corrections which are regularization-scheme-independent. Therefore, corrections that lead to terms like $\Box R$ in $T_{\mu}^{\mu}$ are not considered since their coefficients are regularization-scheme dependent \footnote{ Regularization-scheme dependence of this coefficient has been recently discussed in \cite{alex}}. These terms are also gauge dependent and can be removed by adding a local counterterm.
One could choose working with a specific field theory, e.g., ${\cal N}=4$ Super Yang Mills (SYM) theory since its Trace anomaly is one-loop exact\footnote{Trace anomaly in this theory is one-loop exact since the divergence of $R$-current (R-symmetery is a gauged global symmetry of supersymmetric theories) is in the same supermultiplet of the trace of the stress tensor ${T_\mu}^\mu$ and the divergence of the $R$-current is one-loop exact by Adler-Bardeen Theorem.} \cite{canomaly1,canomaly2,canomaly3,canomaly4}. As a result, no quantum correction is obtained in this theory that modifies Weyl anomaly higher-derivative terms. Although, a full resolution of spacetime singularities needs a theory of quantum gravity, this analysis enables us to investigate how far quantum effects (in matter sources) would go in modifying the model singular behavior.

One of the intriguing consequences of adding these quantum corrections is that the big bang/crunch singularity of the FLRW model changes its nature to a softer type, namely, a sudden singularity. This singularity was introduced in \cite{barrow-s} while studying general features of FLRW cosmologies (a classification of possible future-time singularities was given in \cite{ON-class}). A sudden singularity is characterized by a finite scale factor $a$ and $\dot{a}$, but a divergent $\ddot{a}$. Cosmologists became more interested in sudden singularities after the discovery of cosmic acceleration \cite{cosmos1,cosmos2,cosmos3,cosmos4,cosmos5} since they appear naturally in several cosmological models, see for example \cite{cosmod1,cosmod2,cosmod3,model1,sudden-model1}. One of the important features of a spacetime with a sudden singularity is that its geodesics are well behaved and can be extended across the singular region\cite{jambrina,barrow-g}, therefore, this singularity is traversable. A nice example of this geodesic extension is given by Keresztes et al. \cite{sudden-model1} where the authors considered a cosmological model with anti-Chaplygin gas and dust mixture which suffers from a sudden singularity. Another example for crossing a sudden singularity was considered in \cite{keresztes}.

Weyl anomaly corrections to FLRW models have been considered in the past by a number of authors. In \cite{wald} Wald showed that although density and hubble rate are bounded, field equations cease to admit solutions beyond the singular point. Also, in \cite{hartle} Fischetti et al. showed that the model has a curvature singularity and does not admit any nonsingular solution. Therefore, in later literature this model has been considered singular. Notice that these models are different from that proposed by Starobinsky in \cite{starobinsky} which has no matter content and was shown to admit nonsingular solutions in some rage of the model's parameters. More recently, Weyl anomaly has been discussed in the context of phantom cosmology \cite{ON-phan} and as a quantum escape from sudden singularities in \cite{ON-qes}.

Here we revived interest in this model through showing the following: First, this singularity is weak according to Tipler and Krolak \cite{tipler,krolak}, therefore, it is not a strong physical singularity capable of crushing a finite size object indefinitely. Second, Weyl anomaly corrections changes the nature of the initial singularity from a big bang singularity to a sudden singularity. Third, the two branches of solutions consistent with the semiclassical treatment form disjoint spacetimes. Joining the branches of solutions provides us with a $C^1$ extension to nonspacelike geodesics ending at the singularity. This shows geodesic completeness. Fourth, we use Gauss-Codazzi equations to derive generalized junction conditions for this higher-derivative gravity. The extension of spacetime through joining the two branches of solutions is consistent with these junction conditions and the above geodesic extension.

The rest of the article is organized as follows; in section (2) we introduce Weyl anomaly and consider FLRW model with these higher-derivative terms and its dynamics. In section (3) we show that the singularity of our model is a weak singularity according to both Tipler and Krolak, in addition, we show the criterion for extending geodesics beyond a certain point. In section (4) we show that the singularity is traversable through introducing a $C^1$ extension to nonspacelike geodesics beyond the singular point. We formulate generalized junction conditions for this higher-derivative gravity. We analyze the spacetime formed through joining two branches of solutions of our model using these junction conditions.

\section{FLRW Cosmology with Weyl Anomaly}

\subsection{Weyl Anomaly}
The general form of Weyl anomaly in four dimensions  \cite{Duff} is given by
:

\be
\langle \,T_{\mu}^{\mu}\,\rangle=c_1 E_4+c_2 I_4+c_3 \Box R.
\ee
where ${\rm c}$'s are spin dependent constants, $R$ is Ricci scalar, $E_4$ is the Euler density and $I_4$ is Weyl tensor squared which have the following forms
\bea
E_{4}&=&{1\over 64}\left(R_{\mu\nu\alpha\beta}R^{\mu\nu\alpha\beta}-4R_{\mu\nu}R^{\mu\nu}+R^2\right) \nonumber\\
I_{4}&=&-{1\over 64}\left(R_{\mu\nu\alpha\beta}R^{\mu\nu\alpha\beta}-2R_{\mu\nu}R^{\mu\nu}+{1 \over 3}R^2\right).
\eea
The first term in the trace is known as type``A'', the
next term is type``B'', and the last is type``D", see Ref.\cite{deser}.
It is well known that the Weyl anomaly can not be written as a variation of a local effective action. In addition, the coefficients of all terms are regularization-scheme-independent {\sl except} the type~D anomaly. The last term can be written as a variation of local geometrical terms and therefore, can be removed by a suitable addition of local counterterms. The coefficient of type"D" term is regularization-scheme-dependent, in addition, it is gauge-dependent as well. In general, one can choose a regularization scheme in which this term has a vanishing coefficient which we are going to adopt here.

Here we consider a collection of free conformally invariant fields coupled to a conformally flat metric, $g=\Omega^2\eta$. In this case, the renormalized vacuum expectation value (VeV) of the stress tensor is completely determined by the anomaly up to a local (but not geometrical) traceless conserved tensor $T^{(m)}_{\mu\nu}$ \cite{B-C,BD}, which is going to play the role of a conserved conformal matter source ($T^{(m)}_{\mu\nu}$ is called $^{(4)}H_{\mu\nu}$ in \cite{BD}). The renormalized VeV of Stress-tensor \cite{B-C,BD} has the following form
\bea\langle {T(g)^{(ren)}}_{\mu\nu}\rangle=T^{(m)}_{\mu\nu}+ \alpha
{H_{\mu\nu}}^{(1)}+  \beta {H_{\mu\nu}}^{(3)} ,
\label{stresstwo}
\eea
where, ${H_{\mu\nu}}^{(1)}$ and ${H_{\mu\nu}}^{(3)}$ are given by the expressions
\bea
{H_{\mu\nu}}^{(1)}&=&2R_{;\mu\nu}-2g_{\mu\nu}\Box R-{1\over 2}g_{\mu\nu}R^2+2RR_{\mu\nu}\nonumber\\
{H_{\mu\nu}}^{(3)}&=&{1 \over 12}R^2g_{\mu\nu}-R^{\rho\sigma}R_{\rho\mu\sigma\nu}
\eea
Notice the absence of Weyl tensor contribution since it vanishes for a conformally flat background. The coefficients of $H^{(1)}$ and $H^{(3)}$ are spin depend \cite{BD}, therefore, they are different for scalar, spinor and gauge fields. Taking the trace of both sides one gets
\bea \langle {T^{(ren)}}_{\mu}^{\mu}\rangle={-6\alpha }\, \Box R-  {\beta} \,(\,R_{\mu\nu}R^{\mu\nu}-{1 \over 3}R^2\,)  ,
\eea
The last term on the right hand side is proportional to Euler density (for conformally flat background). In addition, since $\alpha$ is the coefficient of the total derivative term in the anomaly one can either choose a regularization scheme in which $\alpha$ vanishes or can add a local counterterm; namely, $R^2$ with the appropriate coefficient to cancel this term. In the following we are going to consider equation (\ref{stresstwo}) with a vanishing $\alpha$. It is worth mentioning here that this class of regularization schemes (with a vanishing $\Box R$) has been used in other approaches in cosmology. For example, in quantum cosmology one can obtain very similar modified Friedmann equations as in \cite{bravinsky1,bravinsky2,bravinsky3} which produce similar Planck-scale corrections to FLRW cosmology at early times.

\subsection{FLRW Cosmology with Weyl Anomaly}
Here we assume that the universe contains a number of free conformally invariant fields at early times. The field theory considered here is generic, but as we mentioned before, one might prefer to work with a specific theory like ${\cal N}=4$ SYM theory since its Weyl anomaly is one-loop exact \cite{canomaly1,canomaly2,canomaly3,canomaly4}. As a result, no corrections to these higher derivative terms that modify the above field equations. The field theory is coupled to the spatially flat FLRW background
 \be ds^2= -dt^2+a(t)^2[\,dx^2+dy^2+dz^2\,].\ee
Since $\alpha$ is set to vanish, the only modification to Einstein field equations is coming from the Euler density, then the field equation reads
\be R_{\mu\nu}-{1 \over 2} g_{\mu\nu}R-\kappa \,\beta\,\left[{R^2 \over 12}g_{\mu\nu}-R^{\rho\sigma}R_{\rho\mu\sigma\nu}\right] =\kappa T_{\mu\nu}^{(m)} \label{HDT}.\ee
where $\kappa =8\pi G$. The coefficient $\beta$ is known to be $\beta=-{1 \over 2880\pi^2}(n_s+11\,n_f+62\,n_v)$ \cite{20ywa}, where $n_s$ is the number of scalars, $n_f$ is the number of Dirac fermions and $n_v$ is the number of vector fields. It is interesting to notice that $\beta$ is always negative for a generic theory that contains scalar, fermion and vector fields. It is intriguing to notice that this fact is crucial for obtaining a maximum energy density and Hubble rate in the FLRW cosmology.

Homogeneity and isotropy requires the components of the matter stress tensor to be ${{T^{(m)}}_{0}}^{0}=-\rho$, \,\,\,${{T^{(m)}}_{1}}^{1}={{T^{(m)}}_{2}}^{2}={{T^{(m)}}_{3}}^{3}=P$, but since it is traceless, we have $P={1 \over 3}\rho$. I.e., the set up is only consistent with radiation equation of state, i.e., conformal matter source. As in standard cosmology, we get two independent equations
\bea \kappa \rho - 3[1+\kappa \beta H^2]\,H^2 =0 \label{FE}\eea
\bea 2 \ddot{a} a(1+2\kappa \beta H^2)+a^2H^2(1-\kappa \beta H^2)+\kappa \, a^2 P=0, \eea
where, $H= {\dot{a}/a}$ is the Hubble rate. The last equation is equivalent to the continuity equation
\be \dot{\rho}+3H(\rho+P)=0. \ee
This last equation can take the following form
\be \dot{H}(1+2\kappa \, \beta H^2)=-{\kappa \over 2}(\rho+P)\label{doth}, \ee
which can be put in a simpler form after using the above equation of state
\be \dot{H}=-2\,\left({1+\kappa\, \beta H^2 \over 1+2\kappa \, \beta H^2}\right) H^2\label{doth}. \ee
As one can notice, equation (8) is a modification of Friedmann equation, which relates $H$ and $\rho$ through the following roots \cite{wald}
\be H= \pm \sqrt{{-1\pm \sqrt{1+{4 \over 3}\beta {\kappa}^{2}\rho} \over 2 \beta {\kappa}} }\label{sol-H}. \ee
Since $\beta<0$, the two roots which describe a universe with $H \leq (2\kappa |\beta|)^{-1/2}$ correspond to two different solutions
\be H_{\pm}= \pm \,{1 \over \sqrt{2|\beta|\kappa}}\,\sqrt{1- \sqrt{1-{4 \over 3}|\beta| {\kappa}^2\rho}}  \label{sol-H}, \ee
For these solutions the Hubble rate is always less than or equal the Planck energy scale and the density have a maximum value $\rho_{max}\sim {\rho}_{planck}$ since $\beta<0$. Now using the continuity equation we get
\be \rho(a)=c\, a^{-4} \label{density-a},\ee
where $c>0$. Using this density in equation (\ref{sol-H}) and rescale $a\rightarrow \eta$, and the time $t \rightarrow \tau$, we express the Hubble rate as
\be h(\tau)={\eta' \over \eta}=\pm\sqrt{1\mp \sqrt{1-\eta^{-4}}\,}\ee
where $t=\tau \,\sqrt{2\beta \kappa}$, \,$a/a_c=\eta$, $a_c^4=4\beta\kappa^2  \,c /3$, and a prime," $'$ " stands for a derivative with respect to $\tau$. Notice that at $\eta=1$, $h=\pm 1$.
\begin{figure}
\centerline{\includegraphics[angle=-90,width=80mm]{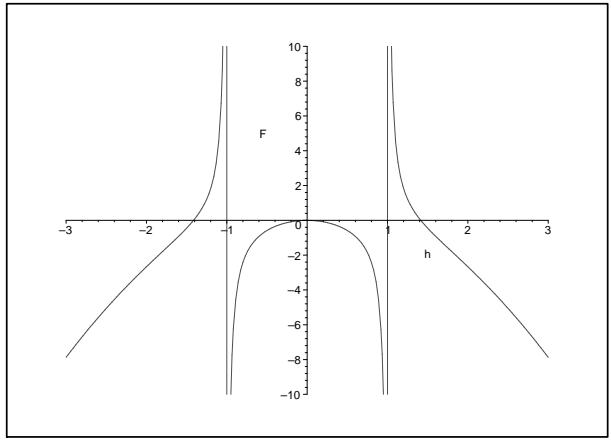}}
\caption{\footnotesize Phase-Space Diagram for the Hubble rate H}
\label{Fig-2}
\end{figure}

Now equation (\ref{doth}) takes the form
\be {h'}=-\left({2-h^2 \over 1-h^2}\right) h^2. \label{phase}\ee
One notices that $h'$ diverges at $h=\pm 1$, this is a curvature singularity since Ricci scalar is proportional to $h'$.

Before we discuss the previous equation let us go back to equation (\ref{FE}). As we have mentioned, equation (\ref{FE}) is a modification of Friedmann equation, which was introduced in \cite{wald}. This relation and its generalizations (including, Cosmological constant, spatial curvature and dark radiation terms) has been introduced and discussed by other relevant approaches in cosmology. Holographic cosmology based on AdS/CFT correspondence \cite{aposto,kiritsis} produces a more general for of this modified Friedmann equation from AdS black holes using holographic renormalization. For recent work in this approach please see \cite{bilic}. Based on quantum cosmological considerations \cite{bravinsky4} the path integral over metrics, where free matter fields are conformally coupled to gravity, is dominated by saddle points which produces the above modified Friedmann equation. Furthermore, thermodynamics at the apparent horizon of a cosmological solution \footnote{One can use thermodynamics at the apparent horizon to produce different corrections to Friedmann equation at Planck scale, which depends on the form of the correction term to Bekenstein-Hawking entropy, for example see \cite{adel+ahmed}} is dictated by the first law which can be used to obtain the above modified Friedmann equation either through using a generalized uncertainty principle \cite{lidsey} or certain volume corrections to Bekenstein-Hawking entropy \cite{viaggui}.

In all these approaches the modified Friedmann equation presented here or its generalizations has been derived and discussed, but issues related to the nature of the singularity, extendability of nonspacelike geodesics and the consistency of that with the field equations through junction conditions were not discussed. In section (3) and (4) we take one step further and analyze these issues. We analyze the nature of the above singularity through studying its strength following Tipler and Krolak. More specifically, we show the possibility of extending nonspacelike geodesics beyond the singular point and use Gauss-Codazzi equations to derive generalized junction conditions for this higher-derivative gravity which are consistent with this geodesic extension.

One might try to use phase-space method (An example of this analysis is in \cite{awad}) to study equation (\ref{phase}) which has a fixed point at $h=0$ (See Figure (1)), but this type of analysis is not adequate to investigate the evolution beyond $h=1$ since $h'$ blows up and the scale factor is discontinues at this point. In fact, we need to check the possibility of extending the spacetime to continue the motion of a test particle across the singular region. In sections (3) and (4) we are going to show that this extension is possible. Before we do that let us have a quick discussion on the mechanical analogue of this cosmology which will clarify few important points on the nature of the above singularity\\

\subsection{ A Mechanical Analogue }

The dynamics of the system can be described by a simple mechanical system that shows some interesting features. Combining the two equations (\ref{sol-H})
and (\ref{density-a}) one gets
\be  E_{\rm kin} + E_{\rm pot} = {1 \over 2} {\eta'}^2-{1 \over 2} \eta^2 ({1\mp\sqrt{1-\eta^{-4}}})=0, \label{energy-cons}\ee
which is the energy conservation equation. In addition, the equation of motion for the mechanical system obtained from (\ref{doth})
\be {\eta}''= \eta {(\sqrt{1-\eta^{-4}} \mp 1) \over \sqrt{1-\eta^{-4}} }. \ee
We have two branches for this mechanical system which give two potential energies and two classes of solutions. The potential energy of the mechanical system is given by
\be V_{\mp}=-{1 \over 2} \eta^2 ({1\mp\sqrt{1-\eta^{-4}}}).\ee

Using equation (\ref{sol-H}) and the above form of Hubble rate one can obtain an exact solution of time $\tau$ as a function of scale factor $\eta$
\be  \tau+c_1 = \pm\left[{\sqrt{2} \over 4} tanh^{-1}(2^{-1/2}\sqrt{1\mp\sqrt{1-\eta^{-4}}}) +{1 \over 2\sqrt{1\mp\sqrt{1-\eta^{-4}}}}\right] \label{soln}.\ee\\
The potential with positive sign corresponds to solutions with $H \geq (2\kappa |\beta|)^{-1/2}$, which is outside the realm of the semiclassical treatment of cosmology since the curvature is greater than or equal $1/l_p^2$. At this high curvature, quantum gravity effects can not be ignored and one should have a full quantum gravity theory to describe this system. Therefore, these solutions are outside the range of validity of our semiclassical treatment. Therefore, we consider only branches with a Hubble rate $H \leq (2\kappa |\beta|)^{-1/2}$.

Notice that although the density and pressure are bounded, the curvature is unbounded at $\eta =1$, since $\eta''$ is unbounded, which are the features of a sudden singularity. This is clear from the expansion of $a$, $H$ and $\dot{H}$ around $t=0$ as we will show in the last section. Anomaly corrections changed the nature of the initial singularity from big bang singularity to a past-sudden singularity. Calculating Ricci scalar one finds\be R \sim {t}^{-{1 \over 2}}.\ee
The mechanical model gives us some intuition on the nature of the singularity which can be summarized as follows;\\

\textbf{i)} As one might notice although the force/acceleration is blowing up at $\tau=0$ and $\eta=1, \eta'=1$, the system needs only a finite amount of work or energy density to evolve from the singular point to any other point with a finite scale factor $\eta$. This is clear from equation (\ref{energy-cons}). In contrast with the usual big bang/crunch singularities which has a different energy conservation equation, namely
\be {1 \over 2} {\eta'}^2-C \eta^{-2} =0,\ee with a singularity at $\eta=0$. In this case, the work needed to take the system away from the singular point is infinite. Compared with the big bang/crunch singularity (i.e., $R \sim {t}^{-2}$), the singularity at hand is much weaker.

\textbf{ii)} From analyzing the mechanical system one realizes that at $\eta=1$ and ${\eta'}=1$, the system can not go to values less that $\eta=1$ and can not stay at this point (${\eta'}=0$), otherwise, it contradicts the energy condition of equation (\ref{energy-cons}). Therefore, one possibility is that the system bounces back with opposite velocity reversing its motion.\\

\textbf{iii)} Using equation(\ref{soln}) with initial condition, $\tau=0$ at $\eta =1$, it is clear that the two regions (for $|h|\leq$ 1) with $\tau \geq 0$ and $\tau<0$ are disjoint, as one can see in Fig.(2). Starting from the region with $\tau<0$ the system can not reach the other side with $\tau \geq 0$, since $\Delta \tau=\tau|_{h=1}-\tau|_{h=-1}\neq 0$. In other words, the spacetime is described by two disjoint manifolds. Therefore, this spacetime can not bounce back with an opposite sign Hubble rate since it is discontinues at the singularity. In the following sections we are going to show that it is possible to glue these two disconnected manifolds (or branches of solutions) to have a bouncing solution which is geodesically complete.
\begin{figure}
\centerline{\includegraphics[angle=-90,width=80mm]{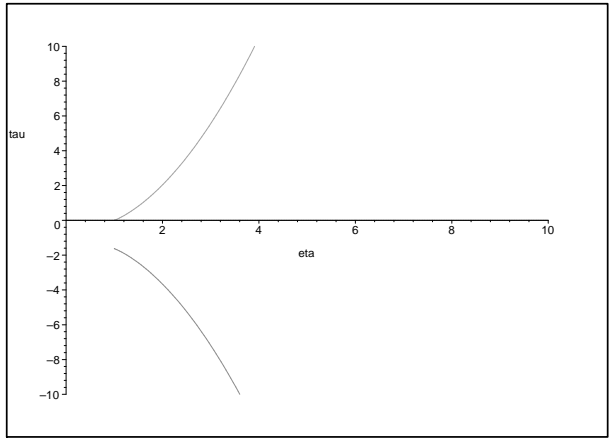}}
\caption{\footnotesize Time $\tau$ as a function of Scale Factor $\eta$ for the two solutions described by $H_-$ and $H_+$, with the condition $\tau(1)=0$. Clearly, the two branches $\tau\geq0$ and $\tau <0$ are disjoint.}\label{Fig-1}

\end{figure}

\section{Geodesic Equations and Strength of Singularity}
In this section we are going to show the criterion for extending geodesics in FLRW models and why this singularity evades Penrose-Hawking singularity theorems. In addition, we check the strength of the singularity and show that it is a weak singularity according to both Tipler and Krolak.

\subsection{Geodesic Equations and Extendability}
Geodesic equations for FLRW metric are giving by
\be {d^2x^i \over d\lambda^2} = 2H\, {dt \over d\lambda}{dx^i \over d\lambda},\ee
\be {dt^2 \over d\lambda^2} = a^2\,H\, \left({dx^i \over d\lambda}\right)^2.\ee
which can be solved to get
\be {dt \over d\lambda} = \pm \sqrt{s+{v^2 \over a^2}}=g(t)\label{t-geod},\ee
\be {dx^i \over d\lambda} = {v^i \over a^2}=f^i(\lambda)\label{x-geod},\ee
where, $s$ and $v^i$ are integration constants and $\lambda$ is a nonspacelike affine parameter. The value of $s$ is related to the length of the tangent vector,$u^a$, where $s=-u^a\,u_a$. Parameter $s$ controls the nature of $\lambda$, if $s$ is $1$, $\lambda$ is a timelike affine parameter but if $s=0$, it is a null affine parameter.\\
According to Picard-Lindel\"{o}f theorem, if $f^i$ and $g$ are continues in $\lambda$ and Lipshitz continues in $t$, there exist unique solutions for the geodesic equations. But as we have seen in the previous section from equation (\ref{soln}), the scale factor is not continues at the singularity, $\tau=0$, therefore, geodesics ends at this point due to the fact that the whole manifold consists of two disjoint parts. Notice that, this does not mean that we can not extend geodesics by defining a new region of spacetime beyond this point. As we will see in the next section it is possible to do this extension since the scale factor and Hubble rate are finite at the singular point, otherwise, this could be impossible as in the big bang/crunch singularity. To understand the role played by the finiteness of the scale factor and Hubble rate at $\tau=0$, let us have a brief discussion on the applicability of Penrose and Hawking singularity theorems to our case.\\

An important defining feature of spacetime singularities is their geodesic incompletness. A spacetime is singular at some region if nonspacelike geodesics end at a finite value of the affine parameter in this region. One of the general results concerning spacetime singularities is the well known singularity theorems of Penrose and Hawking. These theorems prove geodesic inextendibility of spacetimes with certain energy conditions and global properties. An important ingredient in these theorems is the gravitational focusing of congruence of nonspacelike geodesics caused by curvature, which is described by volume expansion $\theta$. The evolution of expansion $\theta$ is governed by Raychaudhuri's equation which leads to the formation of conjugate points for some energy conditions, e.g., strong energy condition. Existence of conjugate points combined with certain global properties implies the occurrence of singularities. One can show that the necessary and sufficient condition (see for example \cite{waldb}) for a point q to be conjugate to p is that the congruence of nonspacelike geodesics emanating from p must have an expansion $\theta\rightarrow -\infty$ at q. \\

The Raychaudhuri equation for our spacetime is nothing but equation (\ref{doth}) and expansion $\theta\propto -H$ which is bounded. As a result, one can conclude that the spacetime under consideration is not singular a la Penrose and Hawking. As we have mentioned before, the singularity under investigation can be classified as a sudden singularity, which is known to be geodesically extendibile\cite{jambrina,barrow-g}.

\subsection{Strength of Singularities}

An important quantity to calculate is the tidal force near a singularity. The tidal force for FLRW is known to be proportional to the acceleration,
\be R^a_{bcd} u^b\xi^c u^d =c {a'' \over a},\ee therefore, it is divergent at the singularity. Here, $\xi^a$ is a deviation vector and $u={\pa \over \pa t}$ is the tangent vector to the curve parameterized by the timelike affine parameter $t$.

Having infinite tidal forces does not necessarily means that the singularity is strong enough to completely crush an infalling finite size object. Important physical criteria for measuring the strength of a singularity has been introduced by Tipler \cite{tipler} and Krolak\cite{krolak}. Tipler's criterion for a strong singularity can be stated as follows; A singularity is called strong if the volume spanned by three orthonormal Jacobi fields shrunk to a zero size along every nonspacelike geodesics at the singularity. Therefore, if the above volume is shrinking to zero size at the singularity, then it is called strong singularity. On the other hand Krolak's criterion for strong singularity is based on having a negative rate of change for the volume at the singularity. As a result, there is a class of singularities which are Tipler weak but Krolak strong.
A lightlike geodesic reaches a Tipler strong singularity at $\lambda=\lambda_0$ if and only if the double integral
 \be \lim_{\lambda \rightarrow \lambda_0} \int_0^{\lambda} d\lambda'\int_0^{\lambda'}d\lambda'' R_{ab}u^a u^b,\ee
diverges. For Krolak, a lightlike geodesics meets a strong singularity if and only if
 \be \lim_{\lambda \rightarrow \lambda_0} \int_0^{\lambda}d\lambda' R_{ab}u^a u^b,\ee
 diverges.
For the timelike case the necessary and sufficient conditions are a bit different. A timelike geodesic meets a strong singularity according to Tipler if
 \be \lim_{\lambda \rightarrow \lambda_0} \int_0^{\lambda}d\lambda' \int_0^{\lambda'}d\lambda'' R_{ab}u^a u^b,\ee
diverges. For Krolak's definition the condition is
\be \lim_{\lambda \rightarrow \lambda_0} \int_0^{\lambda}d\lambda' R_{ab}u^a u^b,\ee
 diverges. According to Tipler, the necessary condition for a timelike geodesics to meet a strong singularity is that
  \be \lim_{\lambda \rightarrow \lambda_0} \int_0^{\lambda}d\lambda' \int_0^{\lambda'}d\lambda'' |R^i_{ajb}u^a u^b|,\ee
 diverges.
 But according to Krolak, the necessary condition for a timelike geodesics to meet a strong singularity is that
  \be \lim_{\lambda \rightarrow \lambda_0} \int_0^{\lambda} d\lambda' |R^i_{ajb}u^a u^b|,\ee
 diverges.
To calculate the above integrals, one needs to obtain expressions for relevant physical quantities, namely, Hubble rate and scale factor near the singularity (at $t=0$). Using equations (\ref{soln}) or (\ref{doth}) for $t\geq 0$, one gets
\be H(t)=H_0\, [1-(H_0t)^{1/2}] +O(t^2)\ee
and
\be a(t)=a_0\, [1+H_0t]+O(t^2),\ee
where $H(0)=H_0=(2\beta\kappa)^{-1/2}$.\\
Now integrating equation (\ref{t-geod}) for $t(\lambda)$, taking the initial condition $\lambda =0$, at $t=0$, one obtains
\be t(\lambda)= \chi\, \lambda +O({\lambda}^2).\ee
Then,
\be a(\lambda)=a_0 \, [1+\chi\,H_0  \lambda] +O({\lambda}^2),\ee

where, $\chi =(s+{v^2 / a_0^2})^{1 \over 2}$. Calculating the above integrals, which gives necessary and sufficient conditions for timelike and sufficient condition for lightlike cases, one gets
\be \lim_{\lambda \rightarrow 0} \int_0^{\lambda} d\lambda'\int_0^{\lambda'}d\lambda'' R_{ab}u^a u^b= C\,\lim_{\lambda \rightarrow 0}\, \lambda^{3/2}=0,\ee
\be \lim_{\lambda \rightarrow 0} \int_0^{\lambda} d\lambda' R_{ab}u^a u^b= C'\,\lim_{\lambda \rightarrow 0}\, \lambda^{1/2}=0.\ee

Also, calculating the above integrals which gives the necessary condition for timelike cases one gets

\be \lim_{\lambda \rightarrow \lambda_0} \int_0^{\lambda} d\lambda'\int_0^{\lambda'}d\lambda'' |R^i_{ajb}u^a u^b|= C_1\,\lim_{\lambda \rightarrow 0}\, \lambda^{3/2}=0,\ee
 \be \lim_{\lambda \rightarrow \lambda_0} \int_0^{\lambda} d\lambda' |R^i_{ajb}u^a u^b|= C_2\,\lim_{\lambda \rightarrow 0}\, \lambda^{1/2}=0,\ee
which indicates that the singularity is weak a la Krolak and Tipler, therefore, it is not a strong physical singularity capable of crushing a finite size object indefinitely. This results suggests that this spacetime might admit some extension which renders the new spacetime geodesically complete.

\section{Singularity Crossing and Junction Conditions}
In this section we are going to introduce a $C^1$ geodesic extension to nonspacelike geodesics beyond the singularity through joining two branches of solutions. We derive junction conditions for the field equation with Weyl anomaly corrections. We also show that the extended spacetime is consistent with the junction conditions of this higher-derivative gravity.
\subsection{Extending Geodesics Beyond the Singularity}

The geodesics equations for the FLRW metric are given by
\be {dt \over d\lambda} = \pm \sqrt{s+{v^2 \over a^2}}=g(t)\label{t-geod},\ee
\be {dx^i \over d\lambda} = {v^i \over a^2}=f^i(\lambda)\label{x-geod}. \ee
Defining an extended spacetime through joining the two regions with scale factors for $t\geq 0$ and $t<0$, one gets
\be a(t)=a_0\, [1+|H_0t|-{2 \over 3}|H_0t|^{3/2}]+O(t^2).\ee
The above first-order differential equations have unique solutions according to Picard-Lendelof Theorem since $f^i$ and $g$ are Lipshitz continous (i.e., they have a bounded first-derivative).
Integrating equation (\ref{t-geod}), and taking the initial condition $\lambda =0$, at $t=0$, one obtains
\be t(\lambda)= \chi\, \lambda -sign(\lambda){H_0 v^2 \over 2a_0^2}{\lambda}^2+ O({\lambda}^3).\ee
Integrating equation (\ref{x-geod}) and taking the initial condition $\lambda =0$, at $x^i={x_0}^i$, one obtains
\be x^i(\lambda)= {x_0}^i+{v^i \over a_0^2}\, \lambda +sign(\lambda)\,{H_0 v^2 \over a_0^2}\,\chi {\lambda}^2+ O({\lambda}^3).\ee
Notice that these geodesics are $C^1$ and are defined for positive and negative values of $\lambda$. Divergences coming from third-derivatives of $t$ and $x^i$ with respect to affine parameter do not affect geodesics. Therefore, the spacetime defined through these geodesics is complete, since it is defined to all values of the affine parameter $\lambda$ and particles going along them can cross the region at $t=0$ without getting crushed indefinitely. Therefore, the singularity is traversable.

\subsection{Generalized Junction Conditions }
Here we use Gauss-Codazzi equations to derive junction conditions for the higher-derivative gravitational theory given by equation (\ref{HDT}).
Using Gaussian normal coordinates near a hypersurface $\Sigma$ with a metric $\tilde{g}$, the line element takes the form
\be ds^2=\epsilon dw^2+\tilde{g}_{ij}dx^idx^j,\ee
where $n^\mu$ is the normal vector to $\Sigma$, with $n_{\mu} n^{\mu}=\epsilon =-1\, or\, 1$ for spacelike or timelike hypersurface respectively. Extrinsic curvature is given by $K_{ij}=-{1 \over 2} \tilde{g}_{ij,w}$. Components of curvature tensor can be expressed in terms of the curvature tensor of the hypersurface and extrinsic curvature $K_{ij}$ and its derivatives. These are Gauss-Codazzi equations which take the following form

\bea {R^l}_{ijk}&=& {\tilde{R^l}}_{ijk}+\epsilon\, (K_{ij}{K_k}^l-K_{ik}{K^l}_j) \nonumber\\
{R^w}_{ijk}&=&-\epsilon \,(K_{ij|k}-K_{ik|j})\nonumber\\
{R^w}_{iwj}&=&\epsilon \,(K_{ij,w}+K_{il}{K^l}_{k}).
\eea
This leads to the following component for the Ricci tensor
\bea {R^i}_{j}&=& {\tilde{R^i}}_{j}+\epsilon \,({K^i}_{j,w}-K\,{K^i}_j) \nonumber\\
{R^w}_{j}&=&-\epsilon \,({K^i}_ {j|i}-K_{|j})\nonumber\\
{R^w}_{w}&=&\epsilon\, (K_{,w}-tr K^2),
\eea
where $K={K^i}_i$ and $trK^2=K_{ij}K^{ij}.$
One can use the above equations to rewrite the field equations (\ref{HDT}) in terms of hypersurface intrinsic curvatures and $K_{ij}$ and its derivatives. To do that we collect the terms on the left hand side in two terms
\be G_{\mu\nu}+H_{\mu\nu}=\kappa T^{(m)}_{\mu\nu},\ee
where \be H_{\mu\nu}=-\kappa \,\beta\,\left[{R^2 \over 12}g_{\mu\nu}-R^{\rho\sigma}R_{\rho\mu\sigma\nu}\right]. \ee
Let us start with Einstein tensor
\bea {G^w}_w&=& -{1 \over 2} \tilde{R}+{1 \over 2}\,\epsilon\, [K^2-trK^2]\nonumber\\
{G^w}_i&=& -\epsilon \,[{{K_{i}}^m}_{|m}-K_{|i}]\nonumber\\
{G^i}_j&=&  \tilde{G^i}_j+\epsilon\, \left[ ({K^i}_j-{\delta^i}_j\,K)_{,w}-K{K^i}_j+{1 \over 2}{\delta^i}_j K^2+{1 \over 2}{\delta^i}_j \,trK^2 \right].
\eea
Now let us join two spacetimes (i.e., solutions of the field equations) at the hypersurface $w=0$. One observes that to avoid powers of Dirac delta function in curvature tensors one must require the continuity of the metric at the joining surface, i.e., $g_{\mu\nu}(0^+)=g_{\mu\nu}(0^-)$. Now integrating the above components of Einstein tensor one gets
\bea \lim_{\sigma\rightarrow 0}\,\int_{-\sigma}^{\sigma}{G^w}_w\, dw&=& [{{{\textbf{G}}}^w}_w]=0 \nonumber\\
\lim_{\sigma\rightarrow 0}\,\int_{-\sigma}^{\sigma}{G^w}_i\, dw&=& [{{\textbf{G}}^w}_i]=0 \nonumber\\
\lim_{\sigma\rightarrow 0}\,\int_{-\sigma}^{\sigma}{G^i}_j\, dw&=& [{{\textbf{{G}}}^i}_j]=\epsilon\,(\,{{[K^i}}_j]-{\delta^i}_j\,[K]\,) \nonumber\\
\eea
where $[A]=A(0^+)-A(0^-)=A^+ -A^-$, is the jump in "$A$" across the junction at $w=0$. For a moment let us set $H_{\mu\nu}=0$, i.e., considering only GR. In this case, the continuity of ${{\textbf{G}}^w}_w =\kappa\, {\textbf{T}^w}_w$ and ${{\textbf{G}}^w}_i=\kappa\, {\textbf{T}^w}_i$ is a consequence of the continuity of the metric across the junction. But the last equation is telling us that if there is a jump in the extrinsic curvature $K_{ij}$ it will lead to a jump in ${{\textbf{G}}^i}_j=\kappa\, {\textbf{T}^i}_j$, therefore, these components of the stress-energy tensor will have Dirac delta functions. This is a consequence of having terms linear in $K_{ij,w}$ and no higher powers in $K_{ij,w}$ or higher-derivative terms in $w$.

Now let us consider the case with $H_{\mu\nu}\neq 0$ and calculate the following components; ${H^w}_w$, ${H^w}_i$, and ${H^i}_j$ to get the junction condition for the above higher-derivative gravity theory. One find that, ${H^w}_w$ and ${H^i}_j$ contains quadratic expressions in $K_{ij,w}$, but ${H^w}_j$ has only up to linear terms in $K_{ij,w}$. If we naively choose junction condition to be $[K_{ij}]=0$, it will lead to the continuity of all component of $\textbf{T}^{\mu}_{\nu}$ and will not allow any surface layer to form. It is more convenient to split the extrinsic curvature into a trace and traceless part;
\be {\hat{K}}_{ij}= {{K}}_{ij}-{\tilde{g}_{ij} \over 3}\,K.\ee This splitting allows us to deal with ${\hat{K}}_{ij}$ and $K$ separately, which was useful in investigating junction conditions in higher-derivative gravity theory like $F(R)$ theories\cite{sasaki}. Now expressing the above component of $H_{\mu\nu}$ in terms of $K$ and $\hat{K}_{ij}$ instead of $K_{ij}$, we get
\bea {H^w}_w =&&\epsilon\,\beta \, \kappa \hat{K}_{ij,w}\, \tilde{R}^{ij} +\epsilon^2\,\beta \, \kappa \left[ \hat{K}_{ij,w}\,\hat{K}^{ij}_{,w}+\hat{K}^{ij}_{,w}\,\left({4 \over 3}\,K \, \hat{K}_{ij}+{\hat{K}_i}^m \,\hat{K}_{jm}\right)\right. \nonumber\\
&&\left. -\hat{K}_{ij,w}\,\left(3\,K\,\hat{K}^{ij}+2\,\hat{K}^{jr}\,{\hat{K}^i}_r \right)  \right] ,\nonumber\\
{H^w}_i =&&\epsilon^2 \,\beta \, \kappa \left[ \hat{K}_{ij,w}\,\left({\hat{K^{mj}}}_{|m}-{2 \over 3}\,{\hat{K}_{|}}^{j}\right) - {\hat{K}^{rs}}_{,w}\,\hat{K}_{ri|s}- {\hat{K}^{rs}}_{,w}\,\hat{K}_{rs|i}\right],\nonumber\\
{H^i}_j =&& -{1 \over 3} \epsilon\,\beta \, \kappa \left[ K_{,w}\, \tilde{R}\,{\delta^i}_j- K_{,w} \, \tilde{R}^{i}_{j}  -3\,\hat{K^{rs}}_{,w} \, {\tilde{R}^i}_{rjs} \right]+\epsilon^2\, \beta \, \kappa\left[ {K_{,w} \over 3}\left( {3\, \hat{K^i}}_{j,w}\right. \right. \nonumber\\
&&\left. -2\, \hat{K}^{ir}\,\hat{K}_{rj} -{5 \over 3}\, K\,\hat{K^i}_j -{\frac{4}{9}}\,K^2\,{\delta^i}_j \right) -{\hat{K^i}}_{j,w}\left( {2 \over 9}\, K^2+ tr \hat{K}^2 \right) \nonumber\\
&& \left.+ {\hat{K^{rs}}}_{,w} \, \left(
\hat{K_{rj}}\,{\hat{K^i}}_{s}-\hat{K_{rs}}\,{\hat{K^i}}_{j}-{{\delta^i}_j \over 3} K\,\hat{K_{rs}} \right)+ {\hat{K^s}}_{j,w}\,{\hat{K^i}}_{s}\,K \right]. \nonumber\\
 \eea
As one can notice ${H^w}_w$ and ${H^w}_j$ contains quadratic and linear expressions in $\hat{K}_{ij,w}$, no dependence on ${K}_{,w}$, but ${H^i}_j$ depends on ${K}_{,w}$ only linearly, and no higher powers. Therefore, it is convent to have the following junction conditions;
\be [\hat{K}_{ij}] =0, \hspace{0.5 in} [K]\neq 0.\ee
i.e., in addition to the continuity of the metric $\tilde{g}$ at the junction we require the continuity of the traceless part of the extrinsic tensor while allowing for a discontinuity or a jump in its trace. These conditions lead to the following equations;
\bea &&\kappa\, [{\textbf{T}^w}_w]= [{{{\textbf{G}}}^w}_w]+ [\textbf{H}^w_w]=0, \nonumber\\
&&\kappa\, [{\textbf{T}^w}_i]= [{{\textbf{G}}^w}_i]+[\textbf{H}^w_i]=0 , \nonumber\\
&&\kappa\, [{\textbf{T}^i}_j]= [{{\textbf{{G}}}^i}_j]+[\textbf{H}^i_j]=\textbf{S}^i_j
\eea
where,
\be \textbf{S}^i_j =\lim_{\sigma\rightarrow 0}\int_{-\sigma}^{\sigma}dw K_{,w} \left[{ \epsilon \over 3}\left({\beta \kappa}[\tilde{R}^{i}_{j}-\tilde{R} {\delta^i}_j ]-{2}{\delta^i}_j\right) + \epsilon^2 \beta \kappa \left({\hat{K}^i}_{j,w}-{4 \over 27} K^2\, {\delta^i}_j\right)\right], \label{s}\ee
and \be \lim_{\sigma\rightarrow 0}\,\int_{-\sigma}^{\sigma}H^{\mu}_{\nu}dw=[\textbf{H}^{\mu}_{\nu}] .\ee

{\bf Junction Conditions for FLRW with Weyl Anomaly:}\\
We are going to use the extended spacetime
\be a(t)=a_0\, [1+|H_0t|-{2 \over 3}|H_0t|^{3/2}]+O(t^2),\ee
joining the two regions for all values of $t$, to calculate the jumps in the stress tensor components, $[{\textbf{T}^i}_j]$. The Hubble rate and its derivative are given by
\be H(t)=  sgn(t)\,H_0 -sgn(t)\,H_0^{3/2}\,\sqrt{|t|} +O(t). \ee
\be \dot{H}(t)=  2\delta(t)\,H_0 -\delta(t)\,H_0^{3/2}\,\sqrt{|t|}-{H_0^{3/2}\over \sqrt{|t|}}  +O(t). \ee
Our hypersurface is spacelike where $w=t$, $\epsilon=-1$ and $\tilde{g}_{ij}=a(t)^2\, \delta_{ij}$. Now the extrinsic curvature is given by
\be K_{ij}=-{1 \over 2}\,\tilde{g}_{ij,t}=-\dot{a}a\,\delta_{ij},\ee
also, \be \hat{K}_{ij}=0, \hspace{0.5 in} K=-3\,H(t), \hspace{0.5 in} K_{,t}=-3\,\dot{H}(t).\ee
Given the facts that $\tilde{R}^{i}_{j}=0$, since our metric is spatially flat (i.e., $k=0$), and $\hat{K}_{ij}=0$, the expression for $\textbf{S}^i_j$ in equation (\ref{s}), is reduced to
\bea \textbf{S}^i_j&=&\lim_{\sigma\rightarrow 0}\int_{-\sigma}^{\sigma}dt \, K_{,t} \left[{ 2 \over 3} - {4  \beta \kappa \over 27} K^2 \right]\,{\delta^i}_j\nonumber\\
&=&-4\, H_0\,\left[1-{H(0)^2 \over {H_0}^2} \right]\, \delta^i_{j},\eea
where, $H_0 = 1/\sqrt{2 \kappa |\beta|}$. As $t\rightarrow 0$, $H(0) \rightarrow sgn(0)\, H_0$, therefore, it depends on our choice of $sgn(t)$ at $t=0$. We have two choices; either $sgn(0)= \pm 1$, or $sgn(0)\neq \pm 1$. The first choice satisfies the above cosmological equations and surprisingly leaves no Dirac delta functions on the stress tensor i.e., $\textbf{S}^i_j=\kappa\, [{\textbf{T}^i}_j]=0$. One might see that as a natural extension of the spacetime which has no Dirac delta function on the hypersurface at $t=0$. The second choice leads to a Dirac delta function but $H(0)$ will not satisfy cosmological equations. One might think of the last choice as throwing out the singular point and replace it with a regular point which is not constrained by  cosmological equations. We prefer the first choice since it leaves all the values of $H(t)$ constrained by our cosmological equations and more importantly gives no Dirac delta function in pressure terms, therefore, does not change the equation of state.

Notice that the above modified Friedmann's equations and its solutions are completely equivalent to the usual Friedmann's equations of GR with the following equation of state
\be P_e= {\rho_e \over 3}{(1+\rho_e /\rho^0_e) \over (1-\rho_e /\rho^0_e)},\ee
where $\rho^0_e=3 H_0^2$. The effective energy density and effective pressure can be written as
\be \kappa \, \rho_{e}= 3H^2=\kappa (\rho+3\beta H^4)\ee
\be \kappa \, P_{e}= -2\ddot{a}-{\kappa \over 3} \rho_{e}=\kappa (P-3\beta H^4-4\beta H^2\dot{H}).\ee
This interesting equivalence provides us with another way to check the above results of junction conditions for our cosmological model. One can observe that this cosmology has a divergent pressure, $P_e$ as a result of a divergent $\dot{H}$. This is a typical feature of sudden singularities in GR. Now this effective stress tensor satisfies
\be G^{\mu\nu}=\kappa T^{\mu\nu}_e. \ee Now consider the junction condition resulted from joining two solution as before at the hypersurface, $t=0$. We get
\bea &&\kappa\, [{\textbf{T}_e^w}_{w}]= [{{{\textbf{G}}}^w}_w], \nonumber\\
&&\kappa\, [{\textbf{T}_e^w}_{i}]= [{{\textbf{G}}^w}_i]=0 , \nonumber\\
&&\kappa\, [{\textbf{T}_e^i}_{j}]= [{{\textbf{{G}}}^i}_j]=-4\, H_0\, \delta^i_{j},
\eea
Therefore, there is a delta Dirac function in the tress tensor ${T_e^i}_{j}$ components. This is in agreement with the earlier junction analysis since the relation between $P_e$ and $P$ has a linear term in $\dot{H}$.
\section{Conclusion}

It is well known that a full resolution of spacetime singularities requires a consistent theory of quantum gravity. Until we have such a resolution, it is constructive to ask how large is the impact of quantum effects, in matter sources, on the FLRW model singular behavior. Here we have considered the role of Weyl anomaly on modifying FLRW model singular behavior. We only considered anomaly corrections that lead to regularization-scheme and gauge independent corrections. As an example we chose ${\cal N}=4$ Super Yang Mills theory to be our field theory since its Trace anomaly is one-loop exact \cite{canomaly1,canomaly2,canomaly3,canomaly4}. Therefore, no corrections to these higher derivative terms which modify the field equations. Weyl anomaly corrections to FLRW models have been considered in the past, we would like to revive interest in this model through showing; This singularity is weak according to Tipler and Krolak \cite{tipler,krolak}, therefore, it is not a strong physical singularity capable of crushing a finite size object indefinitely. Weyl anomaly corrections changes the nature of the initial singularity from a big bang singularity to a sudden singularity. The two branches of solutions consistent with the semiclassical treatment form disjoint spacetimes. Joining the branches of solutions provides us with a $C^1$ extension to nonspacelike geodesics ending at the singularity, which shows geodesic completeness. We use Gauss-Codazzi equations to derive generalized junction conditions for this higher-derivative gravity. The extension of spacetime through joining the two branches of solutions is consistent with these junction conditions and the above geodesic extension. These results suggests that FLRW cosmology can be described by a contracting phase before going to its expanding phase.\\

{\bf Acknowledgement}\\

I would like to thank A. Shapere, for several discussions and comments.

\end{document}